\documentstyle[prd,aps,epsfig,floats]{revtex}
\begin{document}
\draft
\wideabs{
\title{Overlap Distribution of the Three-Dimensional Ising Model}
\author{Bernd A. Berg$^{1,2}$, Alain Billoire$^2$ and Wolfhard Janke$^3$}
\address{ (E-mails: berg@hep.fsu.edu, billoir@spht.saclay.cea.fr,
wolfhard.janke@itp.uni-leipzig.de)\\
$^1$Department of Physics, The Florida State University, Tallahassee, 
FL 32306, USA\\
$^2$CEA/Saclay, Service de Physique Th\'eorique, 91191 Gif-sur-Yvette, 
France\\
$^3$Institut f\"ur Theoretische Physik, Universit\"at Leipzig,
04109 Leipzig, Germany}
\date{May 15, 2002}
\maketitle
\begin{abstract}
We study the Parisi overlap probability density $P_L(q)$ for the 
three-dimensional Ising ferromagnet by means of Monte Carlo (MC) simulations.
At the critical point $P_L(q)$ is peaked around $q=0$ in contrast with
the double peaked magnetic probability density.
We give particular attention to the tails of the overlap distribution
at the critical point, which we control over up to 500 orders of 
magnitude by using the multi-overlap MC algorithm. 
Below the critical temperature interface tension estimates from the 
overlap probability density are given and their approach to the
infinite volume limit appears to be smoother than for estimates
from the magnetization.
\end{abstract}
}

\narrowtext

\section{Introduction} \label{sec_intro}

In this paper we investigate the two replica overlap probability
density $P_L(q)$ for the three-dimensional ($3d$) Ising model. On a
$L^3$ lattice with periodic boundary conditions $q$ is defined by
\begin{equation} \label{q}
q\ =\ {1\over N}\sum_{i=1}^N s_i^{(1)}\,s_i^{(2)}
~~{\rm with}~~N=L^3\ ,
\end{equation}
where $s_i^{(1)}$ and $s_i^{(2)}$ are the spins of two copies
(replica) of the system at temperature $T=1/\beta$.  The distribution
of the overlap $q$ is of major importance in spin-glass
investigations~\cite{BiYo86,Me87,Fi91,Yo97}, where it plays the role 
of an order parameter, often called {\it Parisi order parameter}.

To our knowledge this quantity has never been investigated for simple 
spin systems like the $3d$ Ising model. One reason is certainly that 
one has in that situation the magnetization $m$ as an explicit order 
parameter at hand and a description of the critical properties based
on the magnetic probability density $P_L^m(m)$ is believed to be
identical to one based on $P_L(q)$, in particular $\langle q\rangle =
\langle m \rangle^2$. However, the overlap probability 
density is an interesting object for study on its own merits and we 
find remarkable differences between the shapes of $P_L(q)$ and
$P_L^m(m)$. Therefore, we find it worthwhile to have the properties
of $P_L(q)$ documented for the Ising model, which is by orders of
magnitude easier to simulate than spin glasses, since the dynamics
is much faster and only one (instead of many) realization needs to
be simulated.

In the vicinity of the critical point,
by finite-size scaling (FSS) arguments~\cite{Fi71} $P_L(q)$ can, in
leading order for $L$ large, be written as
\begin{equation} \label{Pp}
P_L(q)={1\over \sigma_L}\,P'(q')~~{\rm with}~~q'= {q\over \sigma_L}\ .
\end{equation}
Here $P'$ is a universal, $L$-independent function and $\sigma_L$ is
the standard deviation of $q$ with respect to the 
probability density $P_L(q\in[-1,+1])$
(or $P_L(q\in[0,1])$ when appropriate).

A major focus of our investigation is on the tails of the $P_L(q)$
distribution, which we control for $L=36$ at $T_c$ over 500 orders 
of magnitude by using the multi-overlap MC algorithm~\cite{BJ98}. This
is also of interest in view of a conjecture by Bramwell {\em et al.\/} 
\cite{Br00,Br01} that a variant of extreme order statistics
describes the asymptotics of certain probability densities for a
large class of correlated systems. Besides the Ising model with some
$T(L)\to T_c$ as $L\to\infty$, their class includes the $2d\ XY$ model
in the low temperature phase, turbulent flow problems, percolation
models and some self-organized critical phenomena. For large $L$ the
asymptotic behavior is claimed to be described by an $L$-independent
curve, which for the overlap variable would read ($q'\to\infty$)
\begin{equation} \label{gumbel}
P'(q') = C\, \exp \left[\, a\,\left( q'-q'_{\max} -
e^{b\,(q'-q'_{\max})}\right) \right]\ .
\end{equation}
Here $C$, $a$, $b$ are constants and $q'_{\max}={q_{\max}/\sigma_L}$,
where $q_{\max}$ is the position of the maximum of the probability
density $P_L(q)$ at positive $q$. Equation~(\ref{gumbel}) is a
variant of Gumbel's first asymptote~\cite{gumbel},
see~\cite{Gu58,Ga87} for reviews of extreme order statistics.

However, Eq.~(\ref{gumbel}) is in contradiction with the widely 
accepted large-deviation behavior, based on the proportionality
of the entropy with the volume~\cite{Bo87},
\begin{equation} \label{standard}
P_L(q)\ \propto\ \exp \left[ -N\, f(q)\right]\ ,
\end{equation}
where, for large $N$, $f(q)$ does not depend on $N$. Our data support 
Eq.~(\ref{standard}).  Using the multimagnetical 
approach~\cite{BHN93} a similar study of the tails could be performed 
for the magnetic probability density $P_L^m(m)$, but this is outside 
the scope of our present paper.

We like to point out that for the overlap distribution of spin glasses 
the status of Eq.~(\ref{standard}) is unclear due to the quenched average.
Our previously reported result~\cite{BBJ02} demonstrates that for the 
$3d$ Edwards-Anderson Ising spin glass a probability distribution of 
the form~(\ref{gumbel}) gives an excellent description of
the tails of the Parisi overlap distribution. Because of the special
nature of its phase transition, Eq.~(\ref{standard}) may also
be questioned for the $2d\ XY$ model, where the extreme order
asymptotics~(\ref{gumbel}) 
(more precisely a variant of it)
with $a=\pi/2$ is found in the spin wave
approximation~\cite{Br00,Br01}. However, the range of validity of
this perturbative argument is unclear, at least to us. It may be
worthwhile to employ the methods of Ref.~\cite{BHN93}, or those of
the present paper, to perform a careful numerical  investigation
of the the $2d\ XY$ model with respect to these questions.

We have performed simulations at and below the critical (Curie) 
temperature $T_c$ of the Ising model phase transition. We
approximate $T_c$ by the value of Ref.~\cite{PSWW84}
\begin{equation} \label{Tc}
\beta_c\ =\ {1\over T_c}\ =\ 0.221\,654
\end{equation}
and present our results for $T_c$ in Sect.~\ref{sec_Tc}. Besides
addressing the question of the asymptotic behavior of the overlap
distribution, we estimate the critical exponent ratio $2\beta/\nu$
from the FSS behavior of the standard deviation $\sigma_L$. For the
temperatures below the Curie temperature we choose
\begin{equation} \label{T_below}
\beta_1\ =\ 0.232 ~~{\rm and}~~ \beta_2\ =\ 0.3\ .
\end{equation}
For $\beta=0.232$ multimagnetical results are available~\cite{BHN93a},
which determine the probability density $P_L^m(m)$ of the 
magnetization over many orders of magnitude.

Our numerical results were obtained with the spin-glass code of the
investigations of \cite{BJ98,BBJ02,BBJ00} by simply choosing all the exchange
coupling constants to be equal to $+1$. A code which is specialized
to the Ising model would be far more efficient. Therefore, we have
limited our present simulations to small and medium sized lattices.
As the results appear already quite clear, there seems to be no
particularly strong reason to push on towards (much) larger systems.

\section{Results at the critical temperature} \label{sec_Tc}

\begin{figure}[-t] \begin{center}
\epsfig{figure=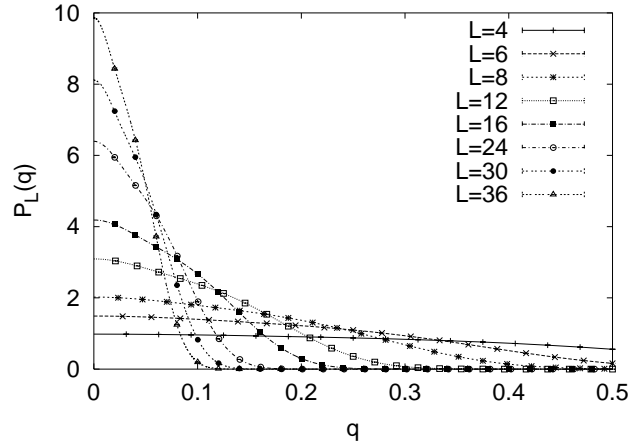,width=\columnwidth} \vspace*{-1mm}
\caption{Overlap probability densities at the critical point
$\beta_c=1/T_c=$ 0.221\,654.} \label{fig_Pq3dITc}
\end{center} \end{figure}

Figure~\ref{fig_Pq3dITc} shows our overlap probability density
results $P_L(q)$ at the critical temperature~(\ref{Tc}). They
rely on a statistics of 32 independent runs (with different pseudo 
random number sequences) for lattices up to size $L=30$ and on
16 independent runs for our largest lattice, $L=36$.  After 
calculating the multi-overlap parameters~\cite{BJ98} the following 
numbers of sweeps were performed per repetition (i.e. independent 
run) : $2^{19}$ for $L=4$, $2^{21}$ for $L=6$, $2^{22}$ for $L=8$, 
$2^{23}$ for $L=12,\ 16$,  $2^{24}$ for $L=24$, $2^{25}$ for
$L=30$ and, again, $2^{24}$ for $L=36$ (with the present computer
program this lattice size became too time consuming to scale
its CPU time properly).
Not to overload Fig.~\ref{fig_Pq3dITc} error bars are only 
shown for selected values of $q$, whereas the lines are drawn
from all data. The probability densities are normalized to
\begin{equation} \label{Pq_norm}
\int_{-1}^{+1} dq\, P_ L(q)\ =\ 1\ ,
\end{equation}
and we show only the $q\ge 0$ part because of the symmetry
$P_L(-q)=P_L(q)$. 
We cut the range at $q=0.5$, because for $L \ge 8$ the
probability densities are almost zero for $q \ge 0.5$.

Somewhat surprisingly we find the maximum of our $P_L(q)$ 
probabilities at $q_{\max}=0$, in contrast to the magnetization
where one finds a double peak at $T_c$, see for instance numerical
data in Ref.~\cite{Bi81} and analytical results of 
Ref.~\cite{BrZi85}, both with periodic boundary conditions.
In our simulation we kept a time 
series for the magnetization, which reproduces the expected 
double peaked histograms at $T_c$ (as the accuracy of our 
magnetization histogram is lower than that of results in the 
literature, we abstain from giving a figure).  Such 
differences are expected. For instance, while the
two low-temperature magnetization values $m=\pm 1$ correspond
to four overlap configurations, two with 
$q=+1$ and two with $q=-1$, the inverse is not true. There are
altogether $2^N$ overlap configurations with $q=+1$ and another
$2^N$ with $q=-1$.

\begin{figure}[-t] \begin{center}
\epsfig{figure=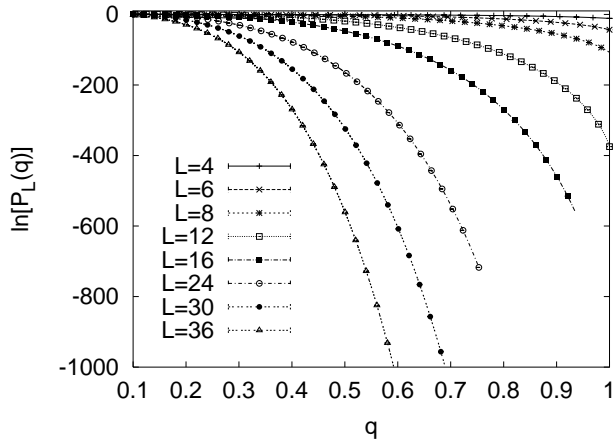,width=\columnwidth} \vspace*{-1mm}
\caption{Logarithm of the overlap probability densities at
$\beta_c=1/T_c= 0.221\,654$.} \label{fig_Pqln3dITc}
\end{center} \end{figure}

Figure~\ref{fig_Pqln3dITc} shows $\ln \left[ P_L(q)\right]$ versus $q$.
The ordinate is cut off at $-1000$, to cover a range with results from
at least two lattice sizes. The $L=36$ lattice continues to exhibit 
accurate results down to $-1200$, thus the data from this system cover
$1200/\ln(10) = 521$ orders of magnitude.

\begin{figure}[-t] \begin{center}
\epsfig{figure=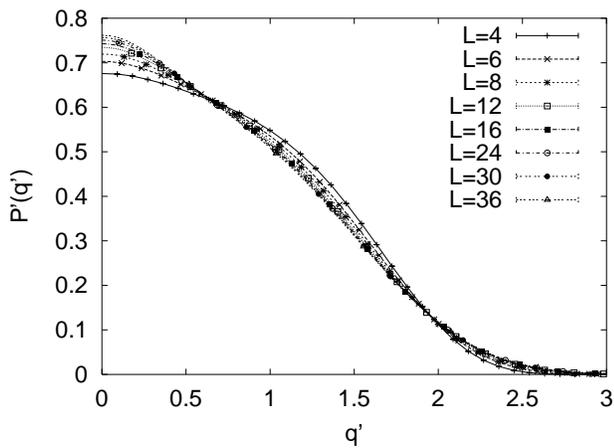,width=\columnwidth} \vspace*{-1mm}
\caption{Rescaled overlap probability densities $P'(q')=$
$\sigma_L\,P_L(q)$ versus $q'=q/\sigma_l$ at the 
critical point.} \label{fig_S3dITc}
\end{center} \end{figure}

The collapse of the $P_L(q)$ functions (\ref{Pp}) on one universal
curve $P'(q')$ is depicted in Fig.~\ref{fig_S3dITc}. The figure shows 
some scaling violations, which become rather small from $L\ge 24$
onwards. The standard deviation $\sigma_L$ behaves with $L$
according to
\begin{equation} \label{q_beta_nu}
 \sigma_L\ \propto\ L^{-2\beta/\nu}\ \left( 1 + 
 c_2\, L^{-\omega} +\dots \right)
\end{equation}
Note that the ratio $\beta/\nu$ is defined for the magnetization,
and by FSS theory~\cite{Fi71} $\sigma_L^m\propto L^{-\beta/\nu}$
holds for the standard deviation of the magnetization. The factor of two
difference in the exponent of Eq.~(\ref{q_beta_nu}) comes from
dimensionality. Scaling relations and estimates of the Ising model
critical exponents are reviewed in Ref.~\cite{PV01}. In particular,
${2\,\beta/ \nu}=d - 2 + \eta$ holds. Our estimate of $2\beta/\nu$ 
from a four-parameter fit of Eq.~(\ref{q_beta_nu}) to our data for 
the standard deviation $\sigma_L$ is $2\beta/ \nu = 1.0293\,(28)$
with $Q=0.31$ the goodness of fit (see~\cite{NumRec} for the 
definition of $Q$). Restricted to our $L\ge 24$ lattices the more 
stable two-parameter fit to the leading behavior of 
Eq.~(\ref{q_beta_nu}) gives
\begin{equation} \label{beta_nu}
{2\,\beta\over \nu}\ =\ 1.030 \pm 0.005~~{\rm with}~~Q=0.36\ .
\end{equation}
The most accurate estimates of the literature~\cite{PV01}
cluster around $\eta=0.036$ with an error of a few units in the last
digit. Within 
the conventional statistical uncertainties
this is consistent with our
$2\beta/\nu$ values. 
The two-parameter fit becomes quickly inconsistent when the smaller lattices
with $L<24$ are included, with a trend towards smaller values of 
$2\beta/\nu$. Therefore, we conjecture that there will
be a slightly increasing trend when larger lattices should become
available. Because its larger error bar reflects to some extent
systematic uncertainties, we prefer Eq.~(\ref{beta_nu}) over
the four-parameter fit as our final estimate.

\begin{figure}[-t] \begin{center}
\epsfig{figure=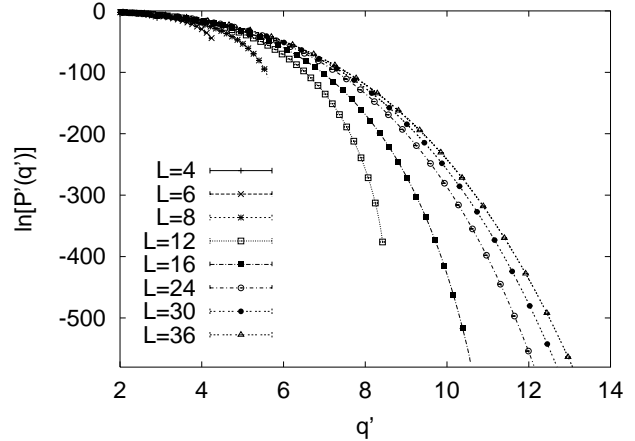,width=\columnwidth} \vspace*{-1mm}
\caption{Logarithm of the rescaled overlap probability densities 
$P'(q')= \sigma_L\,P_L(q)$ versus $q'=q/\sigma_l$ at the 
critical point.} \label{fig_Sln3dITc}
\end{center} \end{figure}

\begin{table}[ht] \begin{center}
\caption{ Deviation points for $\triangle_{36} \ln [P'_L(q')]=0.5$ of 
the $L=4$ to 30 lattices from the $L=36$ result. \label{tab_q_off} }
\vspace*{0.2cm}
\begin{tabular}{|c|c|c|c|c|c|c|c|}       
~~$L$~~ & 4~~~   & 6~~~   & 8~~~   & 12~~~   &
 16~~~  & 24~~~  & 30~~~      \\ \hline
~~$q'$~~& 2.27~~~& 2.46~~~& 2.65~~~& 2.95~~~ &
 3.25~~~& 3.83~~~& 4.54~~~    \\ \hline
~~$q$~~ & 0.781~~& 0.580~~& 0.472~~& 0.351~~~&
 0.288~~& 0.225~~& 0.212~~    \\
\end{tabular} \end{center} \end{table}

In Fig.~\ref{fig_Sln3dITc} we show the logarithm $\ln [P'(q')]$
of the rescaled overlap probability densities and we see a breakdown
of scaling for sufficiently large $q'$. The ordering of the lattices 
is that the rightmost curve corresponds to the $L=36$ lattice. The 
smaller lattices deviate from it. From the left: First, the $L=4$ 
lattice (not visible), next the $L=6$, then $L=8$, $L=12$, $L=16$,
$L=24$ and last $L=30$. 
The agreement is over a larger and larger range in $q'$. 
However, scaled back to $q$, it concerns the vicinity of $q=0$. To
quantify this, we have collected in Table~\ref{tab_q_off} the $q'$ 
and corresponding $q$ values at which the deviation 
$$\triangle_{36}\ln [P'_L(q')] \equiv \ln [P'_{36}(q')]-\ln [P'_L(q')]$$
becomes 1/2, a deviation too small to be visible on the scale of
Fig.~\ref{fig_Sln3dITc}. The $q'$ values are seen to increase,
whereas the corresponding $q$ values decrease.

It is well known that the requirement of consistency of a universal 
probability density~(\ref{Pp}) with the functional form~(\ref{standard})
determines the function $f(q)$. Namely, to leading order the
scaling of the function (\ref{Pp}) $P'_L(q')$ implies that
\begin{equation} \label{gq}
L^d\, f\left( q'\,L^{-2\beta/\nu} \right) = g(q')
\end{equation}
is an $L$-independent function. Therefore,
\begin{equation} \label{scr_fq}
f(q)\ \propto\ q^{d\,\nu/ 2\beta}
\end{equation}
holds. Note that the non-critical Gaussian behavior is a special 
case, obtained for ${d\,\nu/ 2\beta}=2$.

\begin{figure}[-t] \begin{center}
\epsfig{figure=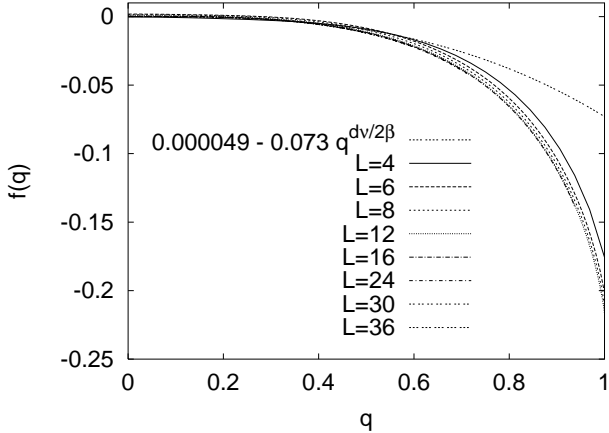,width=\columnwidth} \vspace*{-1mm}
\caption{The functions $f_L(q)$, extracted from
Eq.~(\ref{standard}) for various lattice sizes, are plotted
together with a fit according to Eq.~(\ref{scr_fq}).}
\label{fig_fq3dITc}
\end{center} \end{figure}

\begin{figure}[-t] \begin{center}
\epsfig{figure=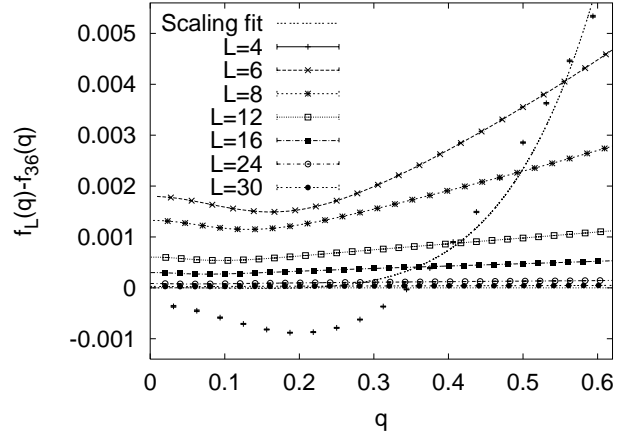,width=\columnwidth} \vspace*{-1mm}
\caption{The functions $f_L(q)$, as well as the fit to
Eq.~(\ref{scr_fq}), with $f_{36}(q)$ subtracted (error bars
are indicated at selected values of $q$). }
\label{fig_fqdel3dITc}
\end{center} \end{figure}

In contrast to Eqs.~(\ref{Pp}) and~(\ref{scr_fq}), the functional
form (\ref{standard}) is expected to hold for all $q$, when $L$ becomes
large. This is easily tested by plotting
\begin{equation} \label{fq}
f_L(q)\ =\ - {1\over N}\, \ln \left[ P_L(q) \right]\ ,
\end{equation}
as is done in Fig.~\ref{fig_fq3dITc}, and seeing if $f_L(q)$ is
$L$-independent up to $O(1/N)$ terms as it should. Not to obscure the 
behavior by too large symbols, the lines are plotted without error bars.
A fit of the scaling form~(\ref{scr_fq}) to our $f_{36}(q)$ data for 
$q<0.2$, $f(q)=0.000049-0.073\,q^{d\nu/2\beta}$ with 
${2\beta/\nu} = 1.030$ from Eq.~(\ref{beta_nu}), is also included in
the figure.

We see excellent convergence towards an $L$-in\-de\-pen\-dent function,
where the higher lying curves correspond to the smaller lattices 
($L=4$ being the one on top). However, the scaling behavior (\ref{scr_fq}) 
only holds in the vicinity of $q=0$.  To make this quantitatively more
precise, we subtract the function $f_{36}(q)$ from the others and 
plot the difference in Fig.~\ref{fig_fqdel3dITc} (at selected
values of $q$ we now include barely visible error bars by plotting
$f_L(q)\pm \triangle f_L(q) - f_{36}(q)$). In the large volume limit
(for both of two lattices) the difference $|f_{L_1}(q)-f_{L_2}(q)|$ 
should be bounded by a constant 
$$ \propto \left( {1\over N_1} - {1\over N_2} \right)\ .$$
We see in Fig.~\ref{fig_fqdel3dITc} that the $L=24$ and 30 curves fall 
almost on the $L=36$ one (the zero line). Note that the figure is cut off at 
$q=0.62$. The number of sweeps needed to propagate the system over the full
admissible $q$-range scales in the multi-overlap ensemble at least
proportional to the system size $N$. Aiming at a comparable statistics
for all system sizes, the required computer time thus grows at least
proportional to $N^2$. Therefore and because of numerical problems with the 
floating point representation caused by the extreme smallness of $P_L(q)$ for 
$q\to 1$ when $L$ is large, we restricted the overlap simulations to 
$q \in [-0.7,+0.7]$ for the lattice sizes $L=16, 24$, and 30, and to
$q \in [-0.62,+0.62]$ for the $L=36$ lattice. 
Nevertheless the smallest 
values of $P_L(q)$ we sampled were those of the $L=36$ lattice.

On the basis of Eq.~(\ref{standard}) the plots of
Figs.~\ref{fig_fq3dITc} and~\ref{fig_fqdel3dITc} had to be
expected. The conjecture (\ref{gumbel}) of Bramwell {\em et al.\/}
\cite{Br00,Br01} appears to be ruled out for the Ising model.
Namely, when Eq.~(\ref{gumbel}) (with $q'_{\max}=0$) and
Eq.~(\ref{fq}) are both valid in some region of 
$bq'=bL^{2\beta/\nu}$, one finds to leading order
\begin{equation} \label{no_gumbel}
b\,q\,L^{2\beta/\nu}\ =\ d\,\ln(L) + \ln [f(q)]\ ,
\end{equation}
and $f(q)$ can only be $L$-independent if $b$ is not a constant, but
depends on $q$ and $L$.

\section{Results below the critical temperature} \label{sec_Tbelow}

\begin{figure}[-t] \begin{center}
\epsfig{figure=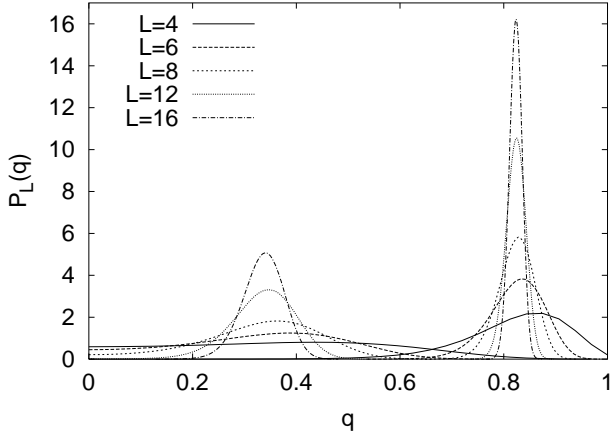,width=\columnwidth} \vspace*{-1mm}
\caption{Overlap probability densities at $\beta=0.232$ (left set
of curves) and $\beta=0.3$ (right set of curves).}
\label{fig_2Pq3dI}
\end{center} \end{figure}

\begin{figure}[-t] \begin{center}
\epsfig{figure=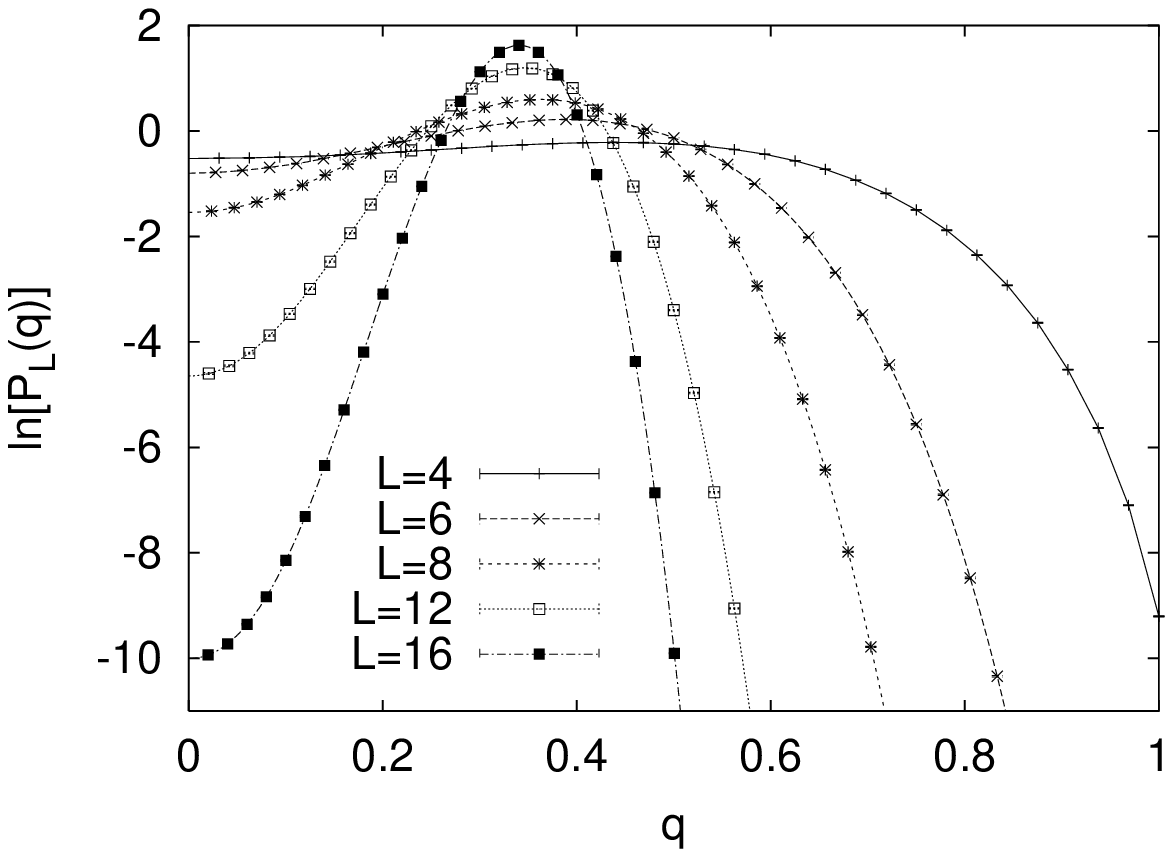,width=\columnwidth} \vspace*{-1mm}
\caption{Logarithms of the overlap probability densities at 
$\beta=0.232$.} \label{fig_Pqln3dITD}
\end{center} \end{figure}

\begin{figure}[-t] \begin{center}
\epsfig{figure=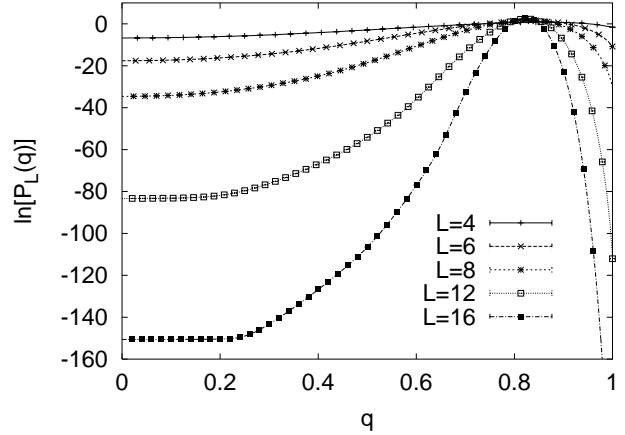,width=\columnwidth} \vspace*{-1mm}
\caption{Logarithms of the overlap probability densities at 
$\beta=0.3$.} \label{fig_Pqln3dITE}
\end{center} \end{figure}

Below the critical temperature we made 16 independent runs per lattice 
size with the following numbers of sweeps per repetition: $2^{16}$ for 
$L=4$, $2^{17}$ for $L=6$, $2^{18}$ for $L=8$, $2^{19}$ for 
$L=12$, $2^{20}$ for $L=16\,(\beta=0.232)$ and $2^{21}$ for 
$L=16\,(\beta=0.3)$.
The overlap probability densities $P_L(q)$ for $\beta=0.232$ and $\beta=0.3$
are shown together in Fig.~\ref{fig_2Pq3dI}. Clearly, the peaks
moved away from zero and are now at $q_{\max}=0.3408$ ($L=16,\,
\beta=0.232$) and $q_{\max}=0.8237$ ($L=16,\, \beta=0.3$). In 
Figs.~\ref{fig_Pqln3dITD} and~\ref{fig_Pqln3dITE} we show
the logarithms of these probability densities at $\beta=0.232$
and $\beta=0.3$, respectively. The scales in these two figures 
are chosen to accommodate all the $P_L(0)$ data, but not
their tails, which continue down to much lower values. For the
largest ($L=16$) difference between $P_L(0)$ and the maximum
of $P_L(q)$, we see that it increases from about four orders
of magnitude at $\beta=0.232$ to about 65 orders of magnitude at
$\beta=0.3$.

\begin{table}[ht]
\begin{center}
\caption[a]{Effective interface tension~(\ref{FsL}) results, $F_{s,L}$,
from the overlap parameter at $\beta=0.232$ and $\beta=0.3$.
At $\beta=0.232$ results for the same quantity obtained 
in Ref.~\cite{BHN93a}
from the magnetization density are included for comparison. \label{tab_FsL} }
\vspace*{0.2cm}
\begin{tabular}{|r|c|c|c|}       
\multicolumn{1}{|c|}{~~$L$~~} & 
$\beta=0.232~~~~$ & $\beta=0.232$~Ref.~\cite{BHN93a}\ 
				   & $\beta=0.3~~~~$  \\ \hline
   4~   & 0.00962 (27)~~~~~ & 0.05297\ (29)~~~~ 
				   & 0.23490\ (33)~~~~\\ \hline
   6~   & 0.01416 (12)~~~~~ & 0.03403\ (15)~~~~
				   & 0.26325\ (20)~~~~\\ \hline 
   8~   & 0.016740 (82)~~~~ & 0.02779\ (13)~~~~
				   & 0.28457\ (20)~~~~\\ \hline
  12~   & 0.020281 (64)~~~~ & 0.02485\ (12)~~~~
				   & 0.29751\ (17)~~~~\\ \hline
  16~   & 0.022715 (34)~~~~ & 0.02521\ (11)~~~~
				   & 0.29959\ (11)~~~~\\
\end{tabular}
\end{center}
\end{table}

Note that the most likely $P_L(0)$ configurations are those where 
one replica stays around magnetization $m=0$ and the other around 
the maximum of the magnetic probability density at positive or 
negative magnetization $m$. It follows that the ratio
\begin{equation} \label{Pratio}
R_L\ =\ { P_L(0)\over P_L(q_{\max}) }\ ,
\end{equation}
where $P_L(q_{\max})$ is the maximum of $P_L(q)$, is related to 
the interface tension $F_s$ according to the formula introduced 
in~\cite{Bi81} for the ratio $P_L^m(0)/P_L^m(q_{\max})$,
\begin{equation} \label{Fs}
R_L = C\, L^p\,\exp\left[-2L^2\,F_s\,\right] + \dots\ .
\end{equation}
Here $C$, $p$ are constants and $p=-1/2$ in the one-loop capillary 
wave approximation\cite{Ca94} or one-loop 
$\Phi^4$-the\-ory~\cite{BrZi85,GeFi90,Mo91}, compare the discussion 
in Ref.~\cite{BNB94}. Two-loop $\Phi^4$-theory 
is considered in Ref.~\cite{HoMu98}. The correction is large 
and it appears that a reliable estimate of $p$ does not exist. 

To determine the interface tension one may first calculate the 
lattice size dependent effective interface tensions
\begin{equation} \label{FsL}
F_{s,L} = - {1\over 2  L^2}\, \ln R_L\ ,
\end{equation}
and then make an extrapolation of $F_{s,L}$ 
for $L\to\infty$. Table~\ref{tab_FsL} collects our $F_{s,L}$ results
where the error bars with respect to the last digits are given in 
parentheses. For the sake of easy comparison, we list also some
$F_{s,L}$ results of Ref.~\cite{BHN93a} at $\beta=0.232$, obtained
by applying the definitions (\ref{Pratio}) and (\ref{FsL}) to the
probability density of the magnetization. It is notable that the
$F_{s,L}$ estimates from the overlap densities increase monotonically
in the listed range of lattice sizes, whereas the $F_{s,L}$ estimates 
from the magnetization show a more complex behavior: Up to $L=12$
they decrease, then they turn around to increase and the increase
has been followed~\cite{BHN93a} up to lattices of size $L=32$.

We pursue a similar fitting strategy as in Ref.~\cite{BHN93a}. As there,
it turns out that our data do not really support fits to more than two
parameters and that including the capillary wave term with the one-loop 
theoretical coefficient $p$ does not lead to any improvements of the 
goodness $Q$ of the fits. In essence we are left with fits to the
leading, likely effective, correction
\begin{equation} \label{FsL_fit}
F_{s,L} = F_s + {a_1\over L}\ ,
\end{equation}
and, due to our small lattice sizes, finite-size corrections are so big 
that the best we can do is a fit of the interface tensions from the 
largest two lattices, $L=12$ and 16. This yields the estimates
\begin{equation} \label{FsTE}
F_s = 0.03002\ (24) ~~{\rm at}~~ \beta = 0.232\ ,
\end{equation}
and
\begin{equation} \label{FsTD}
F_s = 0.30583\ (68) ~~{\rm at}~~ \beta = 0.3\ ,
\end{equation}
which are (under the circumstances of our limited system sizes) 
in reasonably good agreement with results of Hasenbusch and 
Pinn (HP)~\cite{HaPi94}, for a  review see Ref.~\cite{Ha01}.  
Again, our error bars are purely statistical and do not reflect 
systematic errors due to our small lattice sizes.

Our result (\ref{FsTE}) at 
$\beta=0.232$ is lower than the multimagnetical estimate of 
Ref.~\cite{BHN93a}. This moves into the right direction and indicates 
that the resolution of the inconsistency between the multicanonical 
and the HP estimate at $\beta=0.232$, discussed in the paper by 
Zinn and Fisher~\cite{ZiFi96}, has its origin in the 
complex finite-size scaling behavior of $F_{s,L}$ estimates from the 
magnetization, which could be resolved by simulating larger systems. 
It is notable that this difficulty of the extrapolation appears to 
be limited to a small neighborhood of $\beta=0.232$, as the 
multimagnetical $F_s$ estimates~\cite{BHN93a} at $\beta=0.227$ and 
$\beta=0.2439$ are perfectly consistent with HP, see Fig.~1 of 
Ref.~\cite{ZiFi96}.

\section{Summary and Conclusions} \label{sec_conclude}

In Sect.~\ref{sec_Tc} we have analyzed the critical behavior of
the overlap variable $q$. In essence agreement with the standard
scaling picture is found, but with some new insights. In particular,
we exhibit in Table~\ref{tab_q_off} that scaling appears to be 
confined to a small $q$ (but still large $q'=q\,L^{2\beta/\nu}$) 
neighborhood. It may be worthwhile to check whether the magnetic
probability distribution, for which comparable simulations are easier 
to perform, exhibits a similar behavior. Further, we find support
in favor of standard large deviations (\ref{standard}), 
instead of the form (\ref{no_gumbel}) derived from Gumbel's
first asymptote (\ref{gumbel}).

Below the critical point, in Sect.~\ref{sec_Tbelow}, we estimate
interface free energies from our overlap probability densities. The
results are smoother than those from the probability density
of the magnetization~\cite{BHN93a} and tend to reconcile 
discrepancies noted in Ref.~\cite{ZiFi96}. But, like at the critical 
point, considerably larger lattices would be needed to reach high
precision results.

\acknowledgments
We would like to thank Gernot M\"unster und Uwe-Jens Wiese for helpful
correspondence.
This work was in part supported by the US Department of Energy under
contract DE-FG02-97ER41022. The numerical simulations were performed 
on the Compaq SC256 computer of CEA in Grenoble under grant p526
and on the T3E computer of NIC in J\"ulich under grant hmz091.

\end{document}